# Effect of single-particle magnetostriction on the shear modulus of compliant magnetoactive elastomers

Viktor M. Kalita[1,2], Andrei A. Snarskii[1,3], Mikhail Shamonin[4,*], Denis Zorinets[1],

[1]*National Technical University of Ukraine "Kyiv Polytechnic Institute", Prospekt Peremohy 37, 03056 Kiev, Ukraine*
[2]*Institute of Physics NAS of Ukraine, Prospekt Nauky 46, 03028 Kiev, Ukraine*
[3] *Institute for Information Recording NAS of Ukraine, Shpaka Street 2, 03113 Kiev, Ukraine*
[4]*East Bavarian Centre for Intelligent Materials (EBACIM), Ostbayerische Technische Hochschule Regensburg, Prüfeninger Strasse 58, 93049 Regensburg, Germany*

## Abstract

The influence of an external magnetic field on the static shear strain and the effective shear modulus of a magnetoactive elastomer (MAE) is studied theoretically in the framework of a recently introduced approach to the single-particle magnetostriction mechanism [V. M. Kalita *et al*, *Phys. Rev. E* **93**, 062503 (2016)]. The planar problem of magnetostriction in an MAE with soft magnetic inclusions in the form of a thin disk (platelet) having the magnetic anisotropy in the plane of this disk is solved analytically. An external magnetic field acts with torques on magnetic filler particles, creates mechanical stresses in the vicinity of inclusions, induces shear strain and increases the effective shear modulus of these composite materials. It is shown that the largest effect of the magnetic field on the effective shear modulus should be expected in MAEs with soft elastomer matrices, where the shear modulus of the matrix is less than the magnetic anisotropy constant of inclusions. It is derived that the effective shear modulus is non-linearly dependent on the external magnetic field and approaches the saturation value in magnetic fields exceeding the field of particle anisotropy. It is shown that model calculations of the effective shear modulus correspond to a phenomenological definition of effective elastic moduli and magnetoelastic coupling constants. Obtained theoretical results compare well with known experimental data. Determination of effective elastic coefficients in MAEs and their dependence on magnetic field is discussed. The concentration dependence of the effective shear modulus at higher filler concentrations has been estimated using the method of Padé approximants, which correctly predicts that both the absolute and relative changes of the magnetic-field dependent effective shear modulus will significantly increase with the growing concentration of filler particles.

[*] Corresponding author. E-Mail: mikhail.chamonine@oth-regensburg.de

# I. INTRODUCTION

Investigation of magnetoactive elastomers (MAEs) with µm-sized magnetic inclusions (filler particles) is of considerable interest both for industrial applications and fundamental physics of soft matter [1-6]. Under the influence of a magnetic field, MAEs exhibit large magnetostriction (MS) and magnetodeformation which may be several orders of magnitude larger than the magnetostriction of conventional magnetic materials [7 - 12]. The origin of MS in compliant MAEs is different from MS of solid ferromagnets.

In conventional ferromagnetic materials, MS is a consequence of the occurrence of magnetoelastic stresses upon magnetization [13 – 17]. In MAEs, the inherent MS of magnetic inclusions can be neglected, because the MS of these composite materials is much larger than the MS of individual inclusions [18]. MS in MAEs is caused by the action of the magnetized particles on the nonmagnetic matrix. Usually, an elongation of MAE in magnetic fields, which is associated with inter-particle interactions, is investigated [19,20]. For example, in an MAE, MS can be caused by dipole interactions between the magnetic particles, which either attract or repel each other [21-23]. Driven by these forces, filler particles can relatively easily move inside the compliant matrix and deform the sample [24-26]. However, when averaged over the sample, the appearance of MS in MAEs is similar to the MS of conventional ferromagnetic materials.

In this paper, we shall study shear deformations in an MAE in the presence of a magnetic field. This type of deformation is associated with the rotation of magnetic particles by a magnetic field, as it is shown in [18]. Rotation of a particle is a consequence of the action of the torque applied to this particle by a magnetic field. Figuratively speaking, the magnetic field acts as a "screwdriver" on the particle winding up the adjacent elastomer matrix. Investigation of this type of mechanical influence on the MAE by a magnetic field and the description of the impact of local moments of force on the values of shear strain and stress is an emerging statement of the problem in mechanics of composite materials. Transmission of torques from magnetic particles to the polymer matrix and the resulting deformation of a composite material has been previously considered theoretically for MAEs [10,11] and ferrogels [27-29] using alternative material models.

Magnetorheological (MR) or field-stiffening effect is the most notable property of MAEs. It can be defined as the large increase of static or dynamic elastic moduli in externally applied DC magnetic fields. It has been pointed out by several authors that MS and MR effects are inter-related, since their physical origin is the magneto-mechanical coupling between the constituents of the composite materials (see *e.g.* [23,30,31] for recent argumentation). Significant theoretical efforts have been made for disclosing physical mechanisms behind the magneto-mechanical coupling phenomena in MAEs and gels. Macroscopic theories are based on the continuum-mechanical representation of both the elastomer matrix and magnetic inclusions, see *e.g.* [9,10,32-36].

Mesoscopic approaches can account for the granularity and the constitution of magnetic filler particles as separate things, see e.g. [37-44]. An overview of currently available theoretical approaches can be found in Ref. [30]. Alternatively, dynamic properties of MAEs can be phenomenologically modelled using equivalent circuits comprising conventional or generalized springs and dashpots [45-51]. Although significant progress has been achieved in recent years, unified and consistent theoretical description of mechanical behavior of MAEs in magnetic fields is still not complete. Since the underlying physical phenomena are rather complex and different physical effects come simultaneously into play, finding the main physical reason from the results of a numerical experiment could be a daunting task leaving the room for the development of simplified physical models.

The elastic properties of MAEs, as well as of any other composite materials, are characterized by their effective moduli, whose values depend on the elastic characteristics of the matrix and inclusions, the shape of inclusions and their volume concentration [52]. The calculation of the elastic fields in the vicinity of an isolated nonmagnetic inclusion of a regular shape (cylinder, sphere) is a well-known problem [53]. For the pure shear, local deformations of such inclusions are anisotropic and have angular dependence with a period of $\pi$ rad [53].

Recently, a single-particle mechanism of MS in MAEs has been considered by us in two dimensions [18]. It was shown that, in an external magnetic field, a soft magnetic inclusion creates inhomogeneous isotropic local displacements of the elastomer matrix. For the single-particle mechanism of MS [18], local deformations of the elastomer matrix in the vicinity of a magnetic particle differ from local deformations observed under shear strain in the vicinity of a non-magnetic inclusion of the same shape. In the following it will be considered how a particular type of local displacements of the elastomer matrix associated with the single-particle mechanism of MS influences the effective shear modulus of MAEs [53].

The peculiarity of description of mechanical deformations in MAEs, as compared to conventional elastomers, is that the magnetic inclusion has additional degrees of freedom, which must be taken into account when calculating the effective characteristics of MAE in an external magnetic field. As shown in [18], these degrees of freedom are the directions of the magnetic moment vectors of filler particles and the orientations of the easy magnetization axes of magnetic inclusions. Crucial effect of the magnetic anisotropy of filler particles on the equilibrium structure and magnetization of ferrogels has been recently investigated in [54] by coarse-grained molecular dynamics simulations. It has been also shown that the elastic response of ferrogel systems where the particles can be chemically cross-linked into the polymer matrix and the magnetic moments can be fixed to the particles´ axes is strongly influenced by the type of magneto-elastic coupling [55].

In this paper, the effect of magnetic field on the magnitude of the shear modulus of MAEs with a low concentration of magnetically anisotropic particles is studied theoretically. An approach for taking into account the additional degrees of freedom of magnetic filler particles of the composite and calculating the effect of the magnetic field on the elastic moduli is proposed. This procedure allows one to determine the effect of rotation of magnetic particles and their magnetic anisotropy axes on the value of the effective shear modulus of an MAE.

To simplify the mathematical description, the two-dimensional (planar) problem will be solved. To avoid the influence of the demagnetizing field, particles in the form of a thin disk will be considered. For a thin disc, the demagnetization factor vanishes for all directions of the magnetic field in the plane of the disc. For a planar MAE, it can be assumed that an external magnetic field is not distorted and is equal to the internal field which magnetizes the particle. In this case, it is reasonable to assume that the matrix undergoes only planar deformation. These approximations greatly simplify the task of finding required relations between shear strain and stress in magnetic fields. Therefore, these approximations allow one to obtain field dependencies for the shear strain from the region of low fields to the saturation of the shear strain in the framework of a unified approach.

The paper is organized as follows. In Section II our theoretical model is presented and the general expression for the energy of the composite material in a magnetic field is formulated. The linearized case is solved in Section III. Section IV analyzes the results for the specific case of magneto-mechanically soft elastomer matrix. Under magneto-mechanically soft matrix we understand such a matrix where the following condition is fulfilled: $\mu/K \ll 1$, where $\mu$ is the shear modulus of the elastomer matrix and $K$ is the magnetic anisotropy constant of particles. From the point of view of physics the latter condition means that re-orientation of the particle in an external magnetic field can easily deform the matrix. The results are extensively discussed and compared with experimental data in Section V. Conclusions are drawn in the final section.

# II. ENERGY OF A COMPOSITE MATERIAL UNDER SHEAR DEFORMATION IN A MAGNETIC FIELD

We shall study the static simple shearing of an MAE sample (see. Fig. 1). This is the type of shear, as shown in Fig. 1, for which the dynamic shear modulus of MAE samples is determined in low-frequency (oscillation frequency $f \approx$ 1 - 10 Hz) oscillatory shear experiments in the plate-plate configuration with the fixed gap between the plates, see, for example, [47,56-59]. In this shear deformation, the displacement vector $\boldsymbol{u}$ depends only on the $y$-coordinate as $\mathbf{u} = y\psi\mathbf{e}_x$, where $\mathbf{e}_x$ is the unit vector along the $x$-axis. In the polar coordinate system, this vector has two components: $\mathbf{u} = \left(r\psi \sin 2\varphi/2\right)\mathbf{e}_r + r\left((-\psi + \psi \cos 2\varphi)/2\right)\mathbf{e}_\varphi$. The first term in the second bracket is independent of the angle $\varphi$. Because of this term, all points of the sample and, therefore, the entire sample is rotated at an angle $\psi/2$.

Now, if the sample contains a rigid inclusion (*i.e.* the Young's and shear moduli are much greater than those of the elastomer matrix and can be considered to be infinitely large) with circular shape of a radius $r_0$, then, provided that the sample size is much larger than the inclusion's radius, it will also rotate through an angle $\psi/2$ without changing its shape and size. If the inclusion has the anisotropy of magnetic properties, then, in a magnetic field, there will be a torque, acting on the particle as it rotates during the shear deformation. This moment of force will lead to a rotation of the particle and create an additional isotropic [18] deformation around the particle.

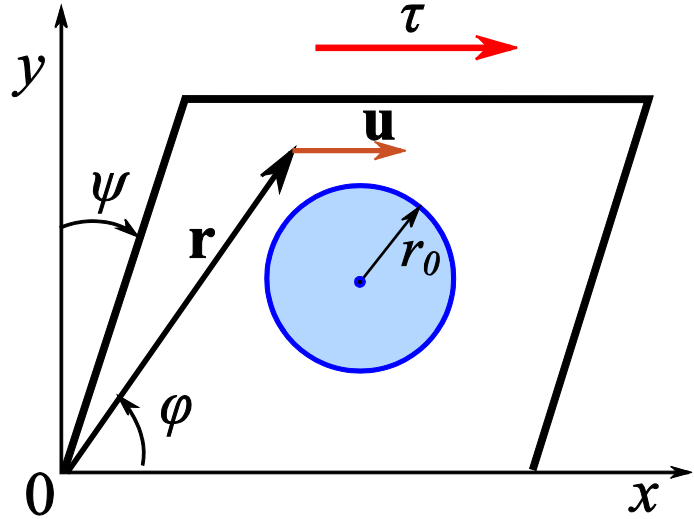

**Fig. 1**. Simple shear test. The engineering shear strain $\psi \ll 1$ is measured from the $y$-axis and considered to be positive in the clockwise direction. The position of a point in the medium is determined by the radius vector **r**. A positive shear stress $\tau > 0$ is shown. The material thickness $D_0$ in the direction perpendicular to the plane of the paper is much smaller than the radius of the inclusion $r_0$: $D_0 \ll r_0$.

Furthermore, we assume that the inclusion has a uniaxial magnetic anisotropy in the *xy*-plane, and that the anisotropy axis is preferable for the direction of the magnetic moment of the particle.

Denote the initial orientation of the particle's easy axis by the angle $\gamma_0$ (see. Fig. 2). Under the shear in the zero magnetic field $\mathbf{H} = 0$, the direction of the particle's easy axis is equal to the angle $\gamma_0 + \psi/2$. If $\mathbf{H} \neq 0$, the angle giving the direction of the particle's easy axis will be equal to $\gamma_0 + \gamma$. Thus, the rotation angle of the particle due to the shear deformation in the magnetic field is equal to the difference of the above said angles: $\gamma_0 + \psi/2 - (\gamma_0 + \gamma) = \psi/2 - \gamma$.

Magnetic energy of the particle and elastic energy of the surrounding matrix taking into account rotation of the particle in a magnetic field [18] can be written as the following sum:

$$E = \left[ -\frac{1}{2} K \cos^2(\gamma_0 + \gamma - \varphi_M) - Hm\cos(\varphi_H - \varphi_M) + 2\mu(\psi/2 - \gamma)^2 \right] V_0, \tag{1}$$

where $V_0$ is the volume of a single particle, $m = M/V_0$ is the particle's magnetization, $K$ is the constant of magnetic anisotropy, $H = |\mathbf{H}|$, $\mu$ is the shear modulus of the matrix, $\varphi_H$ is the angle determining the direction of the magnetic field (cf. Fig 1) and $\varphi_M$ is the angle determining the direction of the magnetic moment ($M = |\mathbf{M}|$) of the inclusion. The first term in (1) describes the energy of the magnetic anisotropy, the second term stands for the Zeeman energy and the third term represents the elastic energy of the matrix taking into account an additional rotation of the particle due to the influence of the magnetic field [18].

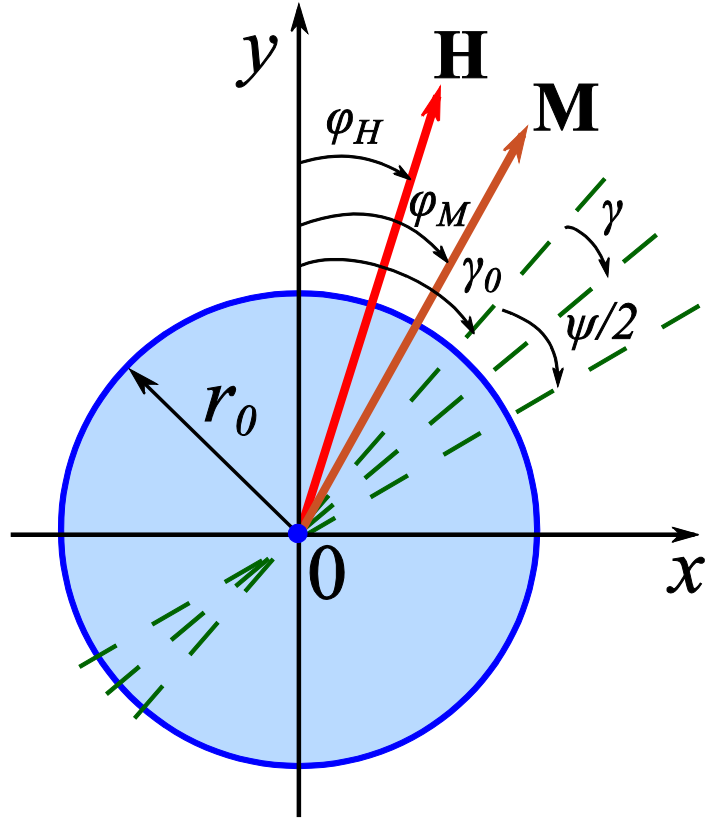

**Fig. 2**. Orientation of the particle's anisotropy axis (dashed line), the magnetic field vector **H** and the magnetic moment vector **M** in the coordinate system shown in Fig. 1. The angles are measured from the y-axis and given positive in the clockwise direction. The angles $\varphi_H$, $\varphi_M$ denote the direction of vectors **M** and **H**, the angle $\gamma_0$ specifies the initial direction of the magnetic anisotropy axis of the particle, $\gamma$ shows the change in the direction of the particle's anisotropy axis after the shear deformation, in which the particle rotates by the angle $\psi/2$.

If all particles are identical and have the same orientation of the anisotropy axis, in the case of low concentration, when the mutual influence of the elastic fields of MAE particles can be neglected, the volume density of single-particle energy in a magnetic field can be written as:

$$\varepsilon_1 = \frac{NE}{V} = p\left[-\frac{1}{2}K\cos^2(\gamma_0+\gamma-\varphi_M) - Hm\cos(\varphi_H-\varphi_M) + 2\mu(\psi/2-\gamma)^2\right], \qquad (2)$$

where $N$ is the number of particles in the volume $V$, $NV_0/V = p$ is the volume fraction of magnetic particles in the composite material, in short their concentration.

To write down the expression for the full energy of an MAE under shear deformation in a magnetic field, the expression (2) should be supplemented by the energy contribution from the shear deformation, $\left(G^{\text{eff}}(0)\psi^2\right)/2$, and the term $\tau\psi$ that takes into account the shear stress. Thus, the final expression for the energy density of a composite with low particle concentration, which includes the magnetic energy of particles, the elastic energy of the matrix caused by particles' rotations in a magnetic field and the elastic energy of the sheared sample under shear stress, takes the form:

$$\mathcal{E} = p\left[-\frac{1}{2}K\cos^2(\gamma_0 + \gamma - \varphi_M) - Hm\cos(\varphi_H - \varphi_M) + 2\mu(\psi/2 - \gamma)^2\right] + \frac{1}{2}G^{\text{eff}}(0)\psi^2 - \tau\psi, \quad (3)$$

where $G^{\text{eff}}(0)$ is the effective shear modulus in the absence of external magnetic field $H=0$ [53,60,61] and $\tau$ is the shear stress.

Thus, for a system of identical particles with the same direction of anisotropy axes, the values of deformation and the angles can be found by minimizing the energy density (3):

$$\frac{\partial \mathcal{E}}{\partial \varphi_M} = p\left[-K\cos(\gamma_0 + \gamma - \varphi_M)\sin(\gamma_0 + \gamma - \varphi_M) - Hm\sin(\varphi_H - \varphi_M)\right] = 0, \quad (4)$$

$$\frac{\partial \mathcal{E}}{\partial \gamma} = p\left[K\cos(\gamma_0 + \gamma - \varphi_M)\sin(\gamma_0 + \gamma - \varphi_M) - 4\mu(\psi/2 - \gamma)\right] = 0, \quad (5)$$

$$\frac{\partial \mathcal{E}}{\partial \psi} = p2\mu(\psi/2 - \gamma) + G^{\text{eff}}(0)\psi - \tau = 0. \quad (6)$$

Equation (4) is obtained by differentiating the energy over the deviation angle of the magnetic moment vector. In (4), the torque caused by the external field on the magnetic moment of the particle is equal to the torque acting on the magnetic moment vector by the anisotropy field $H_A = K/m$, which in the case of the easy-axis anisotropy is directed along the easy magnetization axis of the particle. Equation (5) corresponds to the mechanical equilibrium condition of the particle. It is obtained by differentiation over the rotation angle of the particle. In (5), the torque acting on the particle due to its magnetic moment is equal to the torque caused by the matrix surrounding the particle. Equation (6) is obtained by differentiation over the shear strain. From equation (6) it follows that the magnitude of the shear stress is equal to the sum of the contributions caused by the shear deformation and an additional contribution to the strain created in the matrix by the rotation of particles in a magnetic field.

The following expression can be derived from (4) – (6):

$$\tau = G^{\text{eff}}(0)\psi - \frac{1}{2}pHm\sin(\varphi_H - \varphi_M). \quad (7)$$

From (7) it can be concluded that the torque acting on the magnetic moment vector of a particle results in an additional contribution to the stress.

## III. LINEARIZED PROBLEM

### A. Linearization

Because the system of equations (4) - (6) is nonlinear, it is interesting to consider its solution when all the angles in the expression for the energy (3) are small. We also assume in this section that in the initial state $\tau = 0, H = 0$ and the angle $\gamma_0 \neq 0$. This means that we consider an MAE filled with particles which have the same direction of the anisotropy axis and this axis is not perpendicular to the direction of the applied shear stress (*i.e.*, the *x*-axis, see Fig. 1). For generality, we assume that the magnetic field can be inclined with respect to the shear stress. In this case, the expression for the energy density is simplified and can be written as:

$$e = p\left\{-\frac{1}{2}K\left[1-(\varphi_M - \gamma - \gamma_0)^2\right] - Hm\left[1-\frac{1}{2}(\varphi_H - \varphi_M)^2\right] + 2\mu(\gamma - \psi/2)^2\right\} + \frac{1}{2}G^{\text{eff}}(0)\psi^2 - \tau\psi. \quad (8)$$

Equations of state for the MAE are now linearized:

$$K(\varphi_M - \gamma - \gamma_0) - Hm(\varphi_H - \varphi_M) = 0, \quad (9)$$

$$-K(\varphi_M - \gamma - \gamma_0) + 4\mu(\gamma - \frac{1}{2}\psi) = 0, \quad (10)$$

$$p(-2\mu(\gamma - \frac{1}{2}\psi)) + G^{\text{eff}}(0)\psi - \tau = 0. \quad (11)$$

From the equations of state we get an expression for the shear strain $\psi$, which can be written as

$$\tau = \left[G^{\text{eff}}(0) + \mu p \frac{Hm/K}{(1+Hm/K)(1+4\mu/K)-1}\right]\psi - 2p\mu\frac{\varphi_H - \gamma_0}{(1+Hm/K)(1+4\mu/K)-1}\frac{Hm}{K}. \quad (12)$$

When $H = 0$, equation (12) meets the definition of the effective modulus of the composite material in the absence of a magnetic field: $G^{\text{eff}}(0) = \tau/\psi$.

The product $(\varphi_H - \gamma_0)Hm$ in (12) is equal to the vector product of the magnetization of the inclusion and the magnetic field strength. The right side of Eq. (12) has two contributions. The first contribution is proportional to the strain $\psi$ and independent of the direction of the magnetic field **H**. The second term on the right side of (12) includes angle $\varphi_H$, *i.e.* this term depends on the direction of the magnetic field. At the same time, there are situations when the signs of $\tau$ and $\varphi_H$ may be either the same (*e.g.*, $\tau, \varphi_H > 0$) or different (*e.g.* $\tau > 0$, $\varphi_H < 0$). This, in turn, leads to asymmetry. In the first case, the shear stress $\tau$ and the additional stress created by the action of the magnetic field add and reinforce each other, contributing to the shear strain. In the second case, these stresses counteract each other, reducing the shear strain. This asymmetry of the influence of the oblique magnetic field and the shear is related to the fact that the magnetic field itself is capable of producing shear deformation.

### B. Shear deformation induced by an inclined field

For a system of particles with anisotropy axes perpendicular to the direction of shear stress $\gamma_0 = 0$ and being in a tilted magnetic field, the value of shear strain in the absence of stress $\tau = 0$ is given by

$$\psi = \frac{2\mu K p H m \varphi_H}{4\mu K G^{eff}(0) + Hm\left[G^{eff}(0)(K+4\mu) + Kp\mu\right]}. \tag{13}$$

From (13) it is seen that in weak ($H \to 0$) fields the shear value linearly depends on the field magnitude and it reaches saturation in strong magnetic fields $H \to \infty$.

According to (13), the maximum shear strain induced by an inclined field is achieved in an MAE if $\mu \ll K$. In this case, the easy magnetization axes of the particles are oriented along the magnetic field $\gamma = \psi_H$ and the limiting value of the field-induced shear deformation is equal to:

$$\psi(H \to \infty) = \frac{2\mu}{G^{eff}(0) + p\mu} p\varphi_H \approx 2p\varphi_H. \tag{14}$$

So, the limiting value of the shear deformation induced by a magnetic field in the MAE with $K \gg \mu$, is comparable with the magnitude of the inclination angle of the magnetic field. The magnitude of such a shear deformation is many times greater than the amount of shear induced in the conventional ferromagnetic materials [18].

For MAEs with the small anisotropy constant of particles $K \ll G^{eff}(0)$, the magnitude of shear strain in the limit of a strong inclined field $H \to \infty$ is inclined significantly weakened in comparison to (14):

$$\psi_H(H \to \infty) = \frac{K}{2G^{eff}(0)} p\varphi_H. \tag{15}$$

Thus, the torque created by the magnetic field induces a shear strain in the absence of an external shear stress. The sign of this shear deformation depends on the direction of the applied field.

### C. Shearing in the field orthogonal to the direction of stress

A magnetic field, which is perpendicular to the direction of the external stress (*i.e.* parallel to the *y*-axis in Figure 1), is not capable of inducing the shear strain. Thus, when $\varphi_H = 0$, $\gamma_0 = 0$ and the strain vanishes, $\tau = 0$, the vectors of the particle's magnetic moment and the magnetic field are collinear. Therefore, a perpendicular magnetic field does not generate torque, does not rotate the particle and does not deform the sample.

However, the magnetic field will have significance, if a stress $\tau \neq 0$ is applied to the sample. Under the influence of stress a shear deformation of the sample $\psi \neq 0$ must occur. In this case, the easy magnetization axes will not be directed along the *y*-axis (see Figures 1 and 2) and there will be torques acting on the particles by virtue of the magnetic field. This torque, creating its additional contribution to the stress $\tau$, will affect the amount of the sample's shear by reducing its value and, respectively, increasing the value of the shear modulus.

Indeed, from (12) we obtain the expression:

$$\tau = G^{\text{eff}}(0)\left[1 + \frac{HmK\mu}{G^{\text{eff}}(0)[4\mu K + Hm(K+4\mu)]}p\right]\psi. \tag{16}$$

The effective shear modulus of an MAE in an external magnetic field $H \neq 0$ can be defined as a proportionality factor between the shear strain and stress, so that $G_e(H) = \tau/\psi$:

$$G^{\text{eff}}(H) = G^{\text{eff}}(0)\left[1 + \frac{HmK\mu}{G^{\text{eff}}(0)[4\mu K + Hm(K+4\mu)]}p\right]. \tag{17}$$

Equation (17) describes the change in the effective shear modulus of the MAE under the influence of a magnetic field directed along the anisotropy axis of the particles, and the applied shear stress, which is perpendicular to the magnetic field.

From (12) it is seen that in weak ($H \to 0$) fields, the magnetic-field contribution to the effective shear modulus is proportional to $H$. The magnitude of this additional contribution saturates in large ($H \to \infty$) fields. For the case $K \gg \mu$, we obtain in the large-field limit $H \to \infty$ that

$$G^{\text{eff}}(H \to \infty) = G^{\text{eff}}(0) + p\mu. \tag{18}$$

From (18) it follows that in the saturation limit the magnetic-field contribution to the effective modulus depends on the concentration of the particles and the elasticity of the matrix.

# IV. LIMITING CASES OF MAEs WITH A MAGNETO-MECHANICALLY SOFT MATRIX

Consider some specific cases for the magnetization of the particles possessing strong magnetic anisotropy and located in a compliant matrix. In cases where there is a limiting magnetization (this means that the magnetic moment is directed either along the anisotropy axis or along the external magnetic field), the description of shearing behavior is simplified due to a reduction of the number of degrees of freedom (the number of varying variables in the expression for the energy density (3)).

## A. Highly anisotropic particles

Let, like in the paragraph 1, the axis of the magnetic anisotropy of the particles to deviate from direction the y-axis by an angle $\gamma_0$ (*cf.* Figures 1 and 2). Furthermore, let the magnetic anisotropy of the particles to be so large and the matrix to be mechanically so soft that the condition $K/\mu \gg 1$ is fulfilled. In such an MAE, the magnetic moment vector of the particle can be hardly deflected from the anisotropy axis and it is much easier to rotate the particle. Therefore, the vector of the particle's magnetic moment is always directed along the anisotropy axis of the particle: $\varphi_M = \gamma_0 + \gamma$. When this equality of the angles is fulfilled, the expression for the energy density (8) is simplified and takes the form:

$$\mathcal{E} = p\left[-Hm\cos(\varphi_H - (\gamma_0 + \gamma)) + 2\mu(\psi/2 - \gamma)^2\right] + \frac{1}{2}G^{\text{eff}}(0)\psi^2 - \tau\psi. \tag{19}$$

For particles oriented perpendicular to the direction of the shear stress $\gamma_0 = 0$, we obtain from (19) in the linear case that

$$\psi = \frac{\tau(4\mu + Hm) + 2\mu pHm\varphi_H}{4\mu G^{\text{eff}}(0) + Hm\left[G^{\text{eff}}(0) + p\mu\right]}. \tag{20}$$

It can be easily seen that the expression (20) coincides with (12) if $K \to \infty$. Therefore, the limiting values for the shear magnitude and the shear modulus obtained with the help of (20) will coincide with the values (15) and (18) in saturating magnetic field.

## B. Highly anisotropic particles in saturating magnetic fields

Consider the behavior of the MEA comprising the particles with a large magnetic anisotropy and a soft matrix $\mu \ll K$ in a saturating magnetic field. In a strong magnetic field $H \gg \mu/m$, which in this case may be less than the anisotropy field $H < H_A = K/m$, magnetic moments of particles and anisotropy axes of particles will be oriented along the field: $\varphi_H = \varphi_M$ and $\varphi_H = \gamma_0 + \gamma$.

Therefore, the rotation angle of the particle's magnetization axis will be equal to the difference $\gamma = \varphi_H - \gamma_0$. In this case, the energy density is of the form:

$$\mathcal{E} = p\left\{2\mu[\psi/2 - (\varphi_H - \gamma_0)]^2\right\} + \frac{1}{2}G^{\text{eff}}(0)\psi^2 - \tau\psi. \tag{21}$$

Minimizing (21) with respect to $\psi$, we obtain

$$\psi(H \to \infty) = \frac{\tau + 2\mu p(\varphi_H - \gamma_0)}{\left[G^{\text{eff}}(0) + p\mu\right]}. \tag{22}$$

Equation (22) coincides with the formula (14) for the case $\gamma_0 = 0$ и $\tau = 0$. Thus, the energy density (21) with a smaller number of degrees of freedom gives the result corresponding to the exact solution for the assumed conditions of large anisotropy, softness of the matrix and the saturating magnetic field.

## V. RESULTS AND DISCUSSION

### A. Shearing and effective shear modulus for $\varphi_H = 0$ and $\gamma_0 = 0$

Field dependencies of the shear strain $\psi$ obtained by solving the equations (4) - (6) are shown in Fig. 3. They are plotted for $\varphi_H = 0$ and $\gamma_0 = 0$. In this and the following Figures of Section V, the field is normalized as $h = H/H_A$, where the ratio $H_A = K/m$ is the magnetic anisotropy field of the inclusion. In the calculations, the concentration $p = 0.1$ is considered to be much less than the percolation threshold of composites $p_c \approx 0.5$. If $p = 0.1$, it can be assumed that the effects of elastic interaction between the particles are negligible and the linear approximation is valid. In the absence of a magnetic field at low concentrations, the effective modulus of the composite for a planar elasticity problem with the shear deformation in the direction transverse to the axis of the particles and inclusions of a cylindrical shape as shown in Fig. 1 is described by the approximate expression $G^{\text{eff}}(0) = \mu\left(1 + 2\frac{\lambda + 2\mu}{\lambda + 3\mu}p\right)$, where $\lambda$ and $\mu$ are *Lamé* constants of the matrix [53,61]. It should be noted that the Poisson ratio $v$ of the elastomer matrix is approximately equal to 0.5, from where it follows that $G^{\text{eff}}(0) = \mu(1 + 2p)$. The curves in Fig. 3 are obtained for the shear stress $\tau = 0.001\mu$ and different ratios of the anisotropy constant and the shear modulus of the matrix. The graphs show that the increasing magnetic field counteracts shearing and all the curves decline with the increasing magnetic field. It turns out that the effect of the magnetic field on the magnitude of the shear strain is linear in weak fields and saturates in strong fields. The greatest manifestation of the magnetic field on the magnitude of the shear strain is observed for the magneto-mechanically soft matrix.

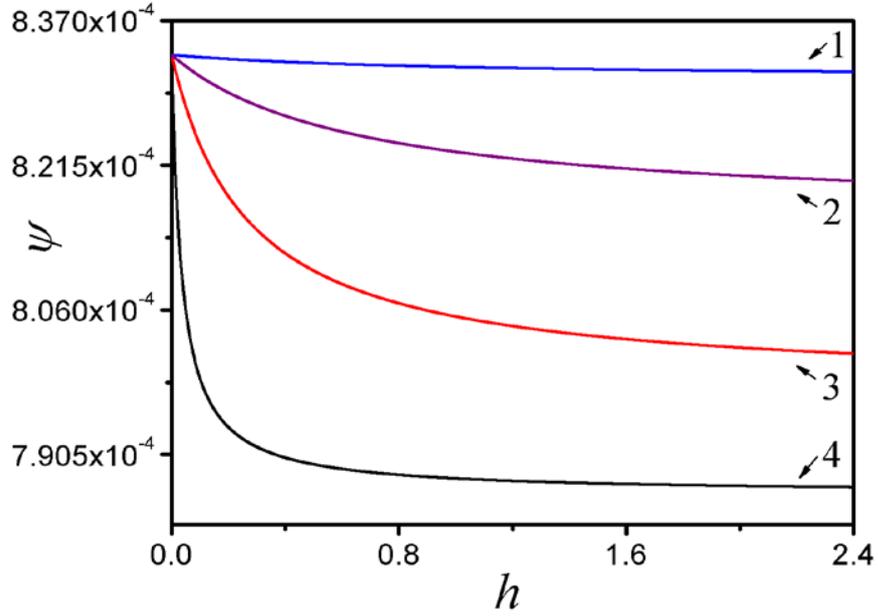

**Fig. 3**. Dependence of the induced shear strain on the magnitude of the normalized magnetic field $h$ for $\varphi_H = 0$ and $\gamma_0 = 0$ and different values of $\mu/K$ = 0.01 (black curve 4), 0.1 (orange curve 3), 0.5 (brown curve 2), 5 (blue curve 1). $\tau = 0.001\mu$.

Fig. 4 shows dependencies of the shear strain $\psi(\varphi_H)$ obtained for different inclinations $\varphi_H$ of the magnetic field $\boldsymbol{H}$. The graphs are calculated in the absence of shear stress $\tau = 0$ at different values of the field and the shear modulus of the matrix. Straight lines on Fig. 4 were obtained by solving the system of equations (9) - (11) for the linear problem, while the curved lines (dash-dotted curves) correspond to the exact solutions of equations (4) - (6). It is seen that the linear approximation provides reasonable solution for a sufficiently broad range of inclination angles. Notable deviations between the solutions occur at larger inclination angles of the magnetic field, where the approximate solution overestimates the magnitude of induced shear strain.

In particular, it is necessary to consider the case of a rigid matrix (curve 1 in Fig. 4). From this curve, it is seen that for $\varphi_H \to \pi/2$ the shear strain vanishes. This is due to the fact that in a rigid matrix the particle practically does not rotate and only the particle's magnetization vector undergoes rotation under the influence of the magnetic field. When there is no shearing and rotation of particles in a rigid matrix, then for $\varphi_H = \pi/2$ in the field which is equal to or greater than the anisotropy field, the angle $\varphi_M \to \pi/2$ (*cf.* [62]).

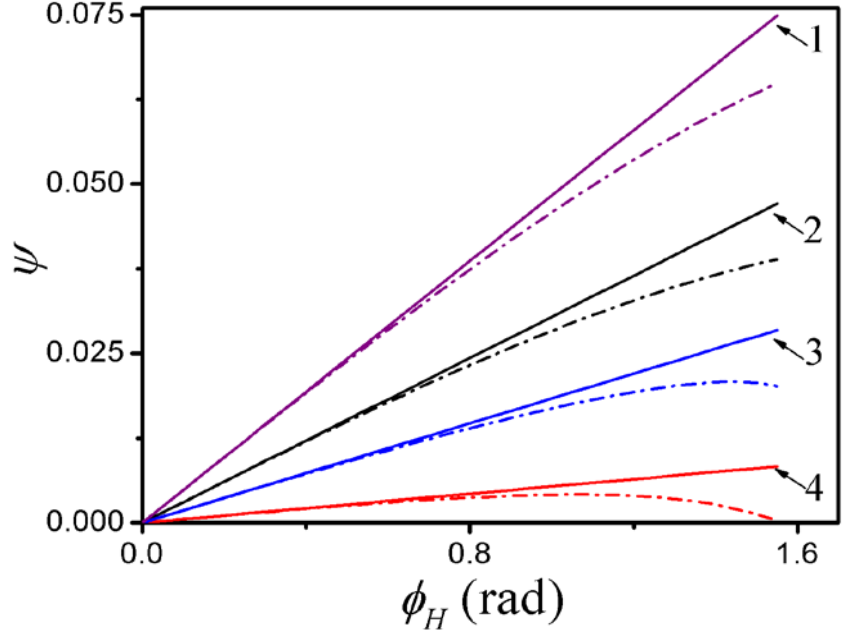

**Fig. 4**. Dependencies of the shear strain of the inclination angle $\varphi_H$ of vector **H**, obtained in the absence of the shear stress $\tau = 0$ for different values of the external magnetic field and the elastic shear modulus of the matrix: $h$=0.035, $\mu/K$=0.035 (1, upper pair of graphs); $h$=0.1, $\mu/K$=0.1 (2, second from the top pair of graphs); $h$=1, $\mu/K$=1 (3, second from the bottom pair of graphs); $h$=2, $\mu/K$=5 (4, bottom pair of graphs). Straight lines were obtained by solving the system of equations (9) - (11) for the linear problem, while the dash-dotted curves correspond to the exact solutions of equations (4) - (6).

Now let us analyze the influence of magnetic field on the shear modulus for the case $\varphi_H = 0$ and $\gamma_0 = 0$. In such an orientation of the magnetic field and the anisotropy axis of the inclusion, there is no shear strain induced by the magnetic field and the value of the shear modulus can be determined from the ratio $\tau/\psi = G^{eff}\left(H, \varphi_H = 0, \gamma_0 = 0\right)$. Using equations (4) - (6), we obtained the shear strains for $\tau = 0.001\mu$, and, consequently, calculated the field dependence of the shear modulus for $H \neq 0$. Fig. 5 shows the dependence of the normalized shear modulus $g^{eff}(H) = G^{eff}(H)/G^{eff}(0)$. In Fig. 5 the graphs for the normalized shear modulus are obtained in Fig. 5 for different ratios of the shear modulus of the matrix and the magnetic anisotropy constant of the inclusion. It is seen that the saturation effect of the shear strain in the magnetic field is accompanied by the saturation of the shear modulus. The dependencies show that the influence of the magnetic field on the effective shear modulus is more pronounced for the magneto-mechanically soft matrix when $\mu/K \ll 1$.

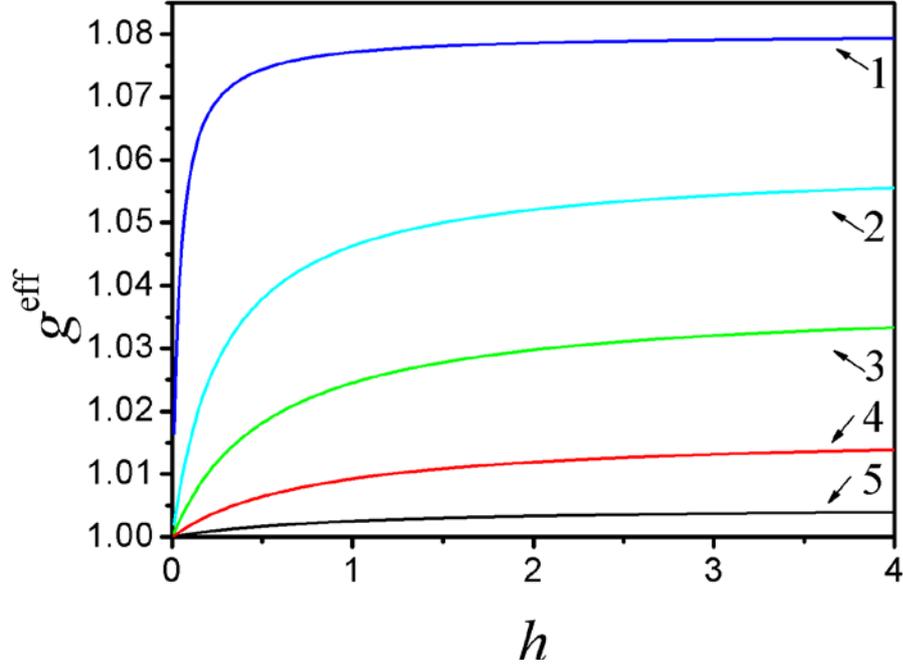

**Fig. 5.** Dependencies of the normalized shear modulus $g^{eff}$ on the normalized magnetic field $h$, obtained for $\tau = 0.001\mu$ at $\varphi_H = 0$ and $\gamma_0 = 0$ and different values of $\mu/K$: $\mu/K = 0.01$ (1); $\mu/K = 0.1$ (2); $\mu/K = 0.3$ (3); $\mu/K = 1$ (4); $\mu/K = 4$ (5).

To illustrate the effect of the elastic properties of the matrix on the magnetic-field enhancement of the shear modulus, the dependencies of the effective shear modulus on the ratio $\mu/K$ of the shear modulus of the matrix and the anisotropy constant have been plotted in Fig. 6 for different constant magnetic fields at $\varphi_H = 0$, $\gamma_0 = 0$, and $\tau = 0.001\mu$. It can be observed that the value of the normalized shear modulus with the increasing matrix rigidity $\mu/K \gg 1$ tends to unity. For small $\mu/K \ll 1$, that is, for the magneto-mechanically soft matrix, the effect of magnetic field on the effective shear modulus is the largest and $g^{eff}$ is equal to its limiting (saturated) value $(g^{eff})_{max} = (1+2p+p)/(1+2p) = 1.083$.

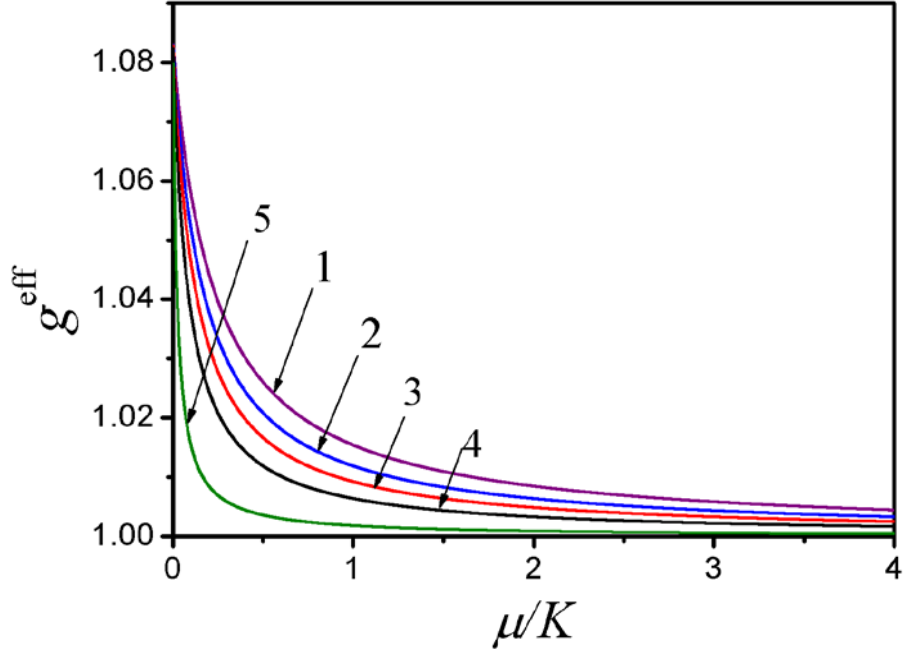

**Fig. 6.** Dependences of the normalized effective shear modulus $g^{eff}$ on the ratio $\mu/K$ of the shear modulus of the matrix and the anisotropy constant of inclusions for different constant magnetic fields at $\varphi_H = 0$, $\gamma_0 = 0$, and $\tau = 0.001\mu$: $h = 10$ (1, the upper curve); $h = 2$ (curve 2, second from the top curve); $h = 1$ (3, middle curve); $h = 0.5$ (4, second from the bottom curve); $h = 0.1$ (5, lower curve).

In Figs. 3 - 6 above, the effect of the magnetic field on the effective shear modulus has been analyzed for $\varphi_H = 0$ and $\gamma_0 = 0$. However, the inclined field induces a shear strain even in the absence of shear stress. Therefore, in an inclined magnetic field, determination of the effective shear modulus requires further discussion.

## B. Effective coefficients in magnetoactive composite materials

The relationship between the mechanical stresses on the specimen (components of stress tensor $\langle \boldsymbol{\sigma} \rangle$, and the deformations of the MAE sample (components of strain tensor $\langle \boldsymbol{\varepsilon} \rangle$) can be written phenomenologically. $\langle ... \rangle$ denotes averaging over the entire volume [63]. A similar approach was used in [62] for crystalline ferromagnetic materials. Besides the terms symmetric with respect to the indices, the elastic stresses in the MAE will also include magnetoelastic coupling caused by the effect of rotation of inclusions in the matrix caused by the torque from an external magnetic field. Given that the torque is proportional to the vector product of the magnetization and the magnetic field strength, the magnetoelastic additional term in the expression for the stress will contain components of an antisymmetric tensor. As a result, we arrive at the following expression for the stress

$$\langle \sigma_{ik} \rangle = C_{iklm}^{\text{eff}} \langle \varepsilon_{lm} \rangle + a_{iklm}^{\text{eff}} \langle m_l \rangle \langle m_k \rangle + b_{iklm}^{\text{eff}} \left( \langle m_l \rangle \langle H_m \rangle - \langle m_m \rangle \langle H_l \rangle \right), \tag{23}$$

where $C_{iklm}^{\text{eff}}$ are effective elastic constants, $a_{iklm}^{\text{eff}}$ and $b_{iklm}^{\text{eff}}$ are effective magnetoelastic coupling coefficients, $\langle m_l \rangle$ and $\langle m_k \rangle$ are projections of the average magnetization vectors, $\langle H_l \rangle$ and $\langle H_m \rangle$ are projections of the magnetic field vector $\mathbf{H} = \langle \mathbf{H} \rangle$, in which the sample is placed. The first magnetoelastic contribution to the stresses (23) is symmetric with respect to the components of the average magnetization, and the second contribution is expressed through the anti-symmetric tensor comprising the projections of the magnetic field strength. Expression (23) is written in the approximation of the smallest exponents with respect to $\langle H_l \rangle$, $\langle m_l \rangle$ and $\langle \varepsilon_{lm} \rangle$. Note that in the linear approximation with respect to $\langle \boldsymbol{\varepsilon} \rangle$, the coefficients $C_{iklm}^{\text{eff}}$ may depend on magnetic field $\mathbf{H}$. In the case of a strongly nonlinear MAE, expansion (23) must contain higher-order terms in the power series and take into account the possibility of the formation in MAE structures, such as particle chain aggregates. General considerations of the interactions between electric, magnetic, and elastic subsystems in nonlinear disordered micropolar media in the framework of phenomenological elastomagnetoelectrostatics can be found e.g. in Ref. [64].

Obviously, the physical approximations made in (23) are satisfied for MAEs with the low particle concentration $p \ll 1$. The approximation made in the formulation of the third term in (23) is also fulfilled if the local additional field is proportional to the average magnetization, what may occur if the particles obey a spatial order in their arrangement [65].

If MAE filler particles have magnetic anisotropy and the particle concentration is low, the third term in (23) may exceed the second term. In this case, for an MAE with the low concentration of ferromagnetic inclusions, the following relation is valid for the shearing experiment of Fig. 1

$$\langle \tau \rangle = G^{\text{eff}}(\mathbf{H})\psi + b^{\text{eff}}\left(\langle m_n \rangle \langle H_t \rangle - \langle m_t \rangle \langle H_n \rangle\right), \tag{24}$$

where $G^{\text{eff}}$ is the effective shear modulus and $b^{\text{eff}}$ is the effective magnetoelastic coupling constant, both depending on $\mathbf{H}$. The index $t$ indicates projection of the vector along the shear stress, and $n$ denotes projection of the vector in the perpendicular direction. In (24), it is taken into account that both components of the magnetic field $\langle H_t \rangle$ and $\langle H_n \rangle$ create a torque on the inclusions, if the magnetization vector is not collinear to $\mathbf{H}$. The expression (24) completely corresponds to the relationship between shear stress and strain in MAEs at small ($H \rightarrow 0$) values of the field, obtained from our model.

From (24) we have that for an MAE with linearly deformable matrix and single-particle mechanism of MS, the magnetic field in the absence of shear strain $\psi = 0$ and $\langle \mathbf{H} \rangle \neq 0$ induces the shear stress

$$\langle \tau(\psi = 0, \mathbf{H}) \rangle = b^{\text{eff}}\left(\langle m_n \rangle \langle H_t \rangle - \langle m_t \rangle \langle H_n \rangle\right). \tag{25}$$

If the shear strain is not equal to zero, the shear stress is equal to

$$\langle \tau(\psi \neq 0, \mathbf{H}) \rangle = G^{\text{eff}}(\mathbf{H})\psi + b^{\text{eff}}\left(\langle m_n \rangle \langle H_t \rangle - \langle m_t \rangle \langle H_n \rangle\right), \tag{26}$$

where $G^{\text{eff}}(\mathbf{H})$ is the effective shear modulus depending on the magnetic field.
From eqs. (25) and (26) we obtain that the effective shar modulus is determined from the following relationship

$$G^{\text{eff}}(\mathbf{H}) = \frac{\tau(\psi \neq 0, \mathbf{H}) - \tau(\psi = 0, \mathbf{H})}{\psi}. \tag{27}$$

Definition of the effective shear modulus (27) is rather general and does not contain the restrictions of the phenomenological description (23), what is confirmed by the calculations of our model.

In MAEs, all effective coefficients may have significant dependences both on the applied magnetic field (see e.g. Fig. 5 and Fig. 6) and on the strain amplitude [66,67]. The definition of (27) can be generalized to other elastic effective coefficients of MAEs by writing the following matrix equation

$$\mathbf{C}^{\text{eff}}\left(\langle\boldsymbol{\varepsilon}\rangle,\langle\mathbf{H}\rangle\right)\langle\boldsymbol{\varepsilon}\rangle = \langle\boldsymbol{\sigma}\rangle\left(\langle\boldsymbol{\varepsilon}\rangle \neq 0, \langle\mathbf{H}\rangle\right) - \langle\boldsymbol{\sigma}\rangle\left(\langle\boldsymbol{\varepsilon}\rangle = 0, \langle\mathbf{H}\rangle\right). \tag{28}$$

Expression (28) allows one to take into account the influence of $\langle\mathbf{H}\rangle$ and $\langle\boldsymbol{\varepsilon}\rangle$ on the value of the effective elastic moduli in MAEs. The numerator in (27) contains two terms, the second of which is the magnetoelastic stress created by a magnetic field in the initially non-deformed sample. If this term is not taken into account, the effective shear modulus determination is incorrect. Obviously, in general, the effective elastic moduli (or compliances) are not the ratio of the corresponding stress and strain components. An analogy with a composite conductive material including thermoelectric effects [68] is outlined in Appendix A.

### C. Determination of the effective shear modulus for $\varphi_H \neq 0$ and $\gamma_0 \neq 0$

To determine the effective shear modulus in an inclined magnetic field, we first calculate $\tau(\psi \neq 0, \mathbf{H} \neq 0)$ and $\tau(\psi = 0, \mathbf{H} \neq 0)$ using equations (4) - (6) and then apply the relationship (27). Fig. 7 shows the dependences of the normalized effective shear modulus $g^{\text{eff}}(\varphi_H)$ on the inclination angle of the magnetic field, which magnitude is kept constant $|\mathbf{H}| = const$. All dependences are plotted for $\mu / K = 1$ and $\gamma_0 = 0$.

From Fig. 7 it is seen that the effective shear modulus depends nonlinearly on the magnetic field direction. The magnitude of the nonlinearity $g^{\text{eff}}(\varphi_H)$ (deviations from straight lines in Fig. 7), was negligible. Note that, in the calculation of $g^{\text{eff}}(\varphi_H)$, the matrix deforms in a linear fashion, as it is implied in (3) and (6). This non-linearity arises from the influence of the magnetic anisotropy. In real MAEs, an alternative to such nonlinearity can be a loss of coupling between the matrix and the particles at significant particle rotations, which are inevitable in soft matrices for large inclination angles of the external field.

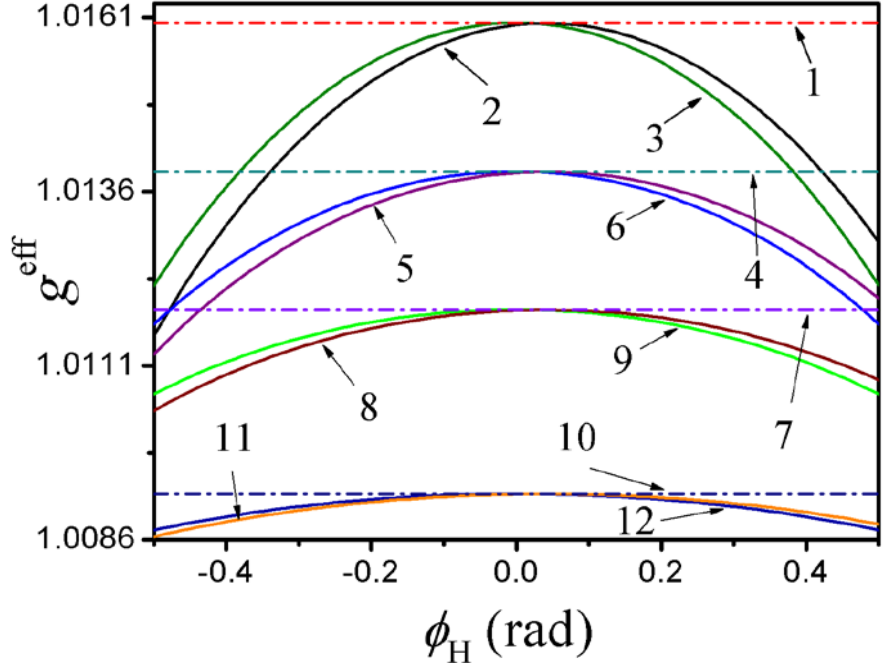

**Fig. 7**. Dependences of the normalized effective shear modulus $g^{eff}(\varphi_H)$ on the inclination angle $\varphi_H$ of the external magnetic field for different values of the normalized magnetic field $h$: $h = 20$ (1 – 3, three upper graphs); $h = 4$ (4 - 6); $h = 2$ (7 - 9); $h = 1$ (10 – 12, three lower graphs). The straight lines 1, 4, 7 and 10 are received from the solution to the linearized problem using equations (9) - (11). The curves 3, 6, 9 and 12 are obtained for the low shear strain $\psi = 0.001$. The curves 2, 5, 8 and 11 are calculated for the large shear strain $\psi = 0.1$. $\mu/K = 1$ and $\gamma_0 = 0$. Notice that curves 3, 6, 9 and 12 appear symmetric with respect to $\varphi_H = 0$ and the maxmum of curves 2, 5, 8 and 11 is shifted towards positive $\varphi_H$.

The curves a1, b1, c1, d1 obtained at low shear strain. If the shear strain is finite (in the calculation of curves a2, b2, c2, d2 the large value $\psi = 0.2$ is assumed), an asymmetry is observed in the dependence $g^{eff}(\varphi_H)$. The maximum of the dependence $g^{eff}(\varphi_H)$ shifts away from the point $\varphi_H = 0$. This shift will be positive if $\psi > 0$ and it will be negative if $\psi < 0$. This leads to the conclusion that when $\psi > 0$ the MAE system in a magnetic field is less rigid for inclinations of the magnetic field in the opposite direction $\varphi_H < 0$ and vice versa.

The magnitude of the normalized effective shear modulus $g^{eff}(\gamma_0)$ depends nonlinearly on the initial orientation angle $\gamma_0$ of the axes of anisotropy inclusion particles. Fig. 8 shows the dependences $g^{eff}(\gamma_0)$ obtained for different constant magnetic fields and $\varphi_H = 0$.

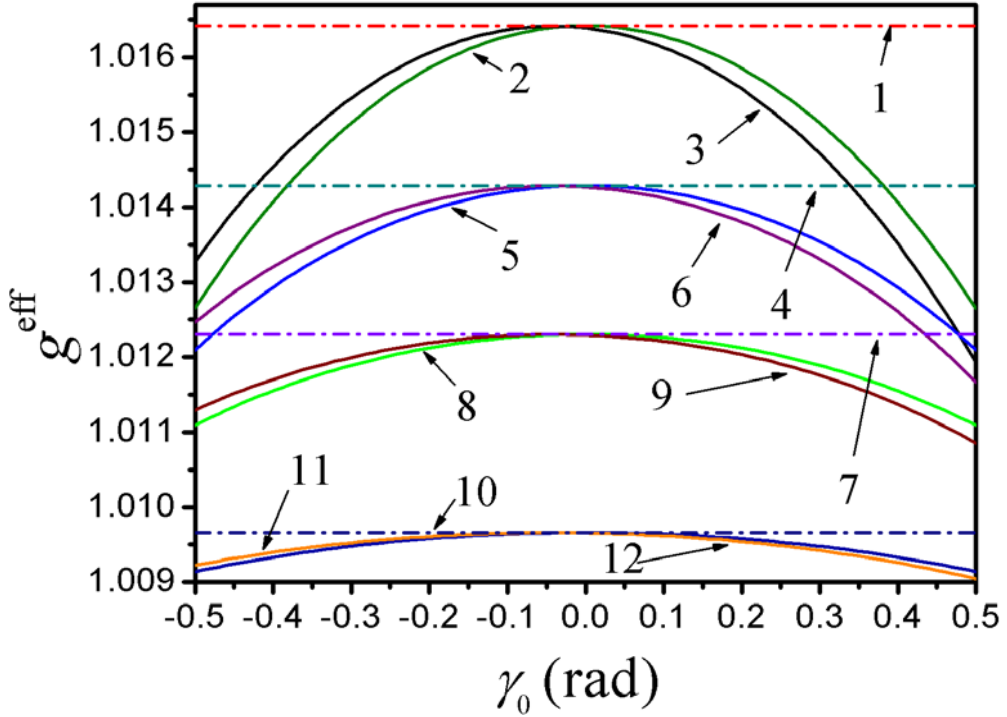

**Fig. 8**. Dependences of the normalized effective shear modulus $g^{eff}(\gamma_0)$ of the angle $\gamma_0$ for the fixed ratio µ / K = 1 and different values of the normalized magnetic field $h$: $h$ = 20 (1 – 3, three upper graphs); $h$ = 4 (4 - 6); $h$ = 2 (7 - 9); $h$ = 1 (10 – 12, three lower graphs). The straight lines 1, 4, 7 and 10 are received from the solution to the linearized problem using equations (9) - (11). The curves 2, 5, 8 and 11 are obtained for the low shear strain $\psi = 0.001$. The curves 3, 6, 9 and 12 are calculated for the large shear strain $\psi = 0.1$. $\mu/K = 1$ and $\varphi_H = 0$. Notice that curves 2, 5, 8 and 11 appear symmetric with respect to $\gamma_0 = 0$ and the maxmum of curves 3, 6, 9 and 12 is shifted towards negative $\gamma_0$.

As can be seen from Fig. 8, there is a weakly expressed nonlinear dependence of the effective shear modulus of the initial orientation of the anisotropy axes of filler particles. For finite values of shear strain dependence, $g^{eff}(\gamma_0)$ is not symmetrical with respect to the sign change of $\gamma_0$.

The observed "anharmonicity" (see Figs. 7 and 8) was not expected by us. It is a manifestation of a strong non-linear behavior of the MAE in a magnetic field. This peculiar effect and variation of $g^{eff}$ with $\gamma_0$ or $\varphi_H$ can be observed only for relatively hard matrices with $\mu/K \geq 1$.

### D. Comparison with experiment

Our simplified model described in the preceding sections refers to a particular realization of the composite material which has not been realized in the experiment yet and might be challenging to achieve in the praxis. The goal of investigating such a theoretical model is that we can identify physical parameters influencing physical effects in more realistic materials. Let us consider as an example the experimental data of Ref. [69] for the shear storage modulus measured as a function of the magnetic flux density $B$ in the range between 0 and 700 mT at a fixed oscillation frequency $f$ of 10 Hz and for different volume fractions of ferromagnetic (iron) filler particles $p$. This experimental set of data was also taken as the reference in previous theoretical works, e.g. [44]. We are interested in the data for the low concentration of filler particles $p = 0.1$ and isotropic samples. Similarly to our model, the shear storages modulus grows with the increasing magnetic field and indicates saturation in large magnetic fields $B > 400$ mT (cf. Figs. 1 and 3 of [69]). This behavior is well known as the MR effect. It has been shown previously in numerous works that the MR effect can be explained by dipolar interactions between ferromagnetic filler particles. May there be a contribution from the single-particle magnetostriction to the observed MR effect? Our model predicts that the maximum contribution from the single-particle magnetostriction to the shear modulus is of the order of magnitude $\Delta G_{max} = \Delta G(B \to \infty) \sim p \cdot G(B=0)$, where $\Delta G(B) = G(B) - G(0)$. It is seen from Fig. 1 of [69] that, for small iron particles (mean particle size about 5 µm), $\Delta G_{max}/G(0) \approx 1.6 - 1.7$, while for large iron particles (mean particle size about 40 µm), $\Delta G_{max}/G(0) \approx 1.3$ (cf. Fig. 3 [69]). It can be concluded that the experimentally observed magnetic-field induced changes of the shear modulus can have a contribution from the single-particle mechanism, since they are of the same order of magnitude and show similar qualitative dependence on the applied magnetic field. The origin of this contribution can be shape deviations of filler particles from the perfect sphere (e.g. ellipsoidal or irregular shape). Indeed, the scanning electron microscope photographs of MAE samples presented in Fig. 4 of Ref. [69] do leave such an impression. The MR effect observed at high concentrations ($p \approx 0.3$) of iron particles was of several orders of magnitude higher $\Delta G_{max}/G(0) \approx 10^2$. In analogy to MR fluids [70], this effect is commonly attributed to rearrangement of filler particles into chain-like aggregates along the magnetic field lines due to magnetic forces acting between them [1,24,71,72]. This simplified physical picture for high concentrations of magnetizable particles has been recently questioned in Ref. [73], where numerical simulations showed that formation of elongated structures becomes impossible due to purely geometrical constraints.

Calculation of the influence of magnetic field on the effective shear modulus of the MAE was performed in the single-particle particle approximation (Maxwell's approximation), where both elastic and magnetic interactions between the particles can be neglected. Note that in spite of the low concentration of particles, they should not be regarded as solitary because the mechanical deformation is self-consistent. Accordingly, the expression for the effective modulus (12) includes only the first degree of concentration. It is well known, both experimentally and theoretically, that the effective properties (for a variety of physical situations), when approaching the concentration of the percolation threshold, are strongly increasing functions and behave like the order parameter in the theory of phase transitions [74,75].

Despite the simplicity of the approach used, it is possible to estimate how the concentration dependence of the effective elastic modulus of MAE will behave in at higher concentrations. To do this, we resort to the method of Padé approximants [68,75]. We write the concentration dependence of the effective modulus which is a first-order polynomial as the ratio of two polynomials. Then, according to (12), taking into account that $G^{\text{eff}}(0) = \mu(1+2p)$, we have

$$G^{\text{eff}}_{\text{Pade}}(H) = \mu \left[ \frac{4\mu + h(K+4\mu)}{8\mu + h(3K+8\mu)} \right] (p_c - p)^{-1}, \tag{29}$$

where

$$p_c = \frac{4\mu + h(K+4\mu)}{8\mu + h(3K+8\mu)}. \tag{30}$$

For $p \ll 1$, the expression $G^{\text{eff}}_{\text{Pade}}(H)$ simplifies, as it should be, into (12).

As follows from (30), in high magnetic fields ($H \gg H_A$), $p_c$ is reduced from ½ for $K \ll \mu$ to ⅓ for $\mu \ll K$. Formula (29) predicts that the magnitude of the magnetic-field dependent effective shear modulus will grow strongly with the increasing particle concentration $p$, what is indeed observed in the experiment. For $H \to 0$, parameter $p_c$ is the well-known percolation threshold, which is somewhat modified in an external magnetic field.

Fig. 9 shows the dependences of the relative magnetic-field induced change of the shear modulus, $MRE = (G^{\text{eff}}(H) - G^{\text{eff}}(0))/G^{\text{eff}}(0)$, on the concentration of filler particles $p$ calculated using Eqs. (29) and (30). It is seen that this relative MR effect (MRE) grows strongly with increasing concentration $p$. In a large magnetic field $H = 10 \cdot H_A$, this effect is pronounced more for magneto-mechanically softer matrices (compare curves 1, 2, 3 with different values of $\mu/K$). At a fixed value of $\mu/K$, the relative change of the shear modulus is increasing with the increasing magnetic field strength $H$ (compare curves 2, 4, 5 with different values of $H/H_A$). These results qualitatively agree with well-known observations from experiments on MAEs. Notice the scale on the vertical axis of Fig. 9. The maximum numbers are about two orders larger than the maximum concentration $p \approx 0.3$. This is the result of using the method of Padé approximants. There will be no such strong growth of MRE within the linear approximation with respect to $p$ (17).

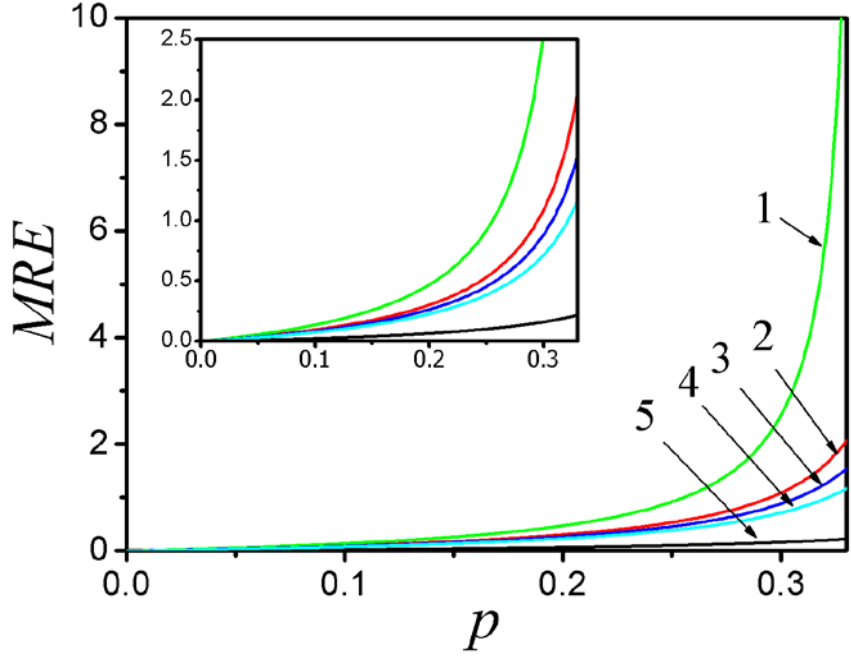

**Fig. 9**. Dependences of the relative magnetic-field induced change of the shear modulus, $MRE = \left(G^{eff}(H) - G^{eff}(0)\right)/G^{eff}(0)$, on the concentration of filler particles $p$ for different ratios of $\mu/K$ and different normalized magnetic fields $h$: (1) − $\mu/K = 1$, $h = 10$; (2) − $\mu/K = 0.1$, $h = 10$; (3) − $\mu/K = 0.01$, $h = 10$; (4) − $\mu/K = 0.1$, $h = 2$; (5) − $\mu/K = 0.1$, $h = 1$. The inset shows the same dependences with the different scale on the vertical axis, where the lines 2 – 5 can be distinguished better. The sequence of curves in the inset is the same as in the main Figure.

## VI. CONCLUSION

In this paper, we have studied the effect of single-particle magnetostriction mechanism on the effective shear modulus of an MAE with a low concentration of ferromagnetic inclusions. The planar (two-dimensional) problem with the inclusions in the form of a disc (platelet) with the axial magnetic anisotropy in the plane of the disc has been solved. It has been shown that any deviation of the magnetic moment of the particle from the easy magnetization axis is accompanied by an enhancement of the effective shear modulus. In particular, the effect of a magnetic field on the effective shear modulus is strongly pronounced in MAS with a magneto-mechanically soft matrix, in which the shear modulus of the matrix is much smaller than the magnetic anisotropy constant. In our model, we were able to demonstrate how the torque of the magnetic field acting on the non-collinear magnetic moment of the particle creates a shear stress. In the phenomenological description, the total torque acting on the particles is determined by the vector product of the particle magnetization in the magnetic field strength and it corresponds to the antisymmetric tensor of the components of these vectors. It is also shown that the effective shear modulus has a strong nonlinear dependence of the external magnetic field characterized by a saturation effect in high magnetic fields much larger than the magnetic anisotropy field. The experimentally observed magnetic-field induced increase of the shear modulus at low filler concentrations $p < 0.1$ can have a significant contribution from the proposed single-particle mechanism. The concentration dependence of the effective shear modulus of MAE at higher filler concentrations has been estimated using the method of Padé approximants, which predicts that the magnitude of the magnetic-field dependent effective shear modulus will significantly grow both absolutely and relatively with the increasing particle concentration $p$.

## APPENDIX A

In the absence of a temperature gradient, an electric field **e** creates an electric current $\mathbf{j} = \sigma \mathbf{e}$. In the case where the material has thermoelectric properties, the thermal gradient creates an additional electric current and the second term appears in the current density $\mathbf{j} = \sigma \mathbf{e} + \sigma \alpha (-gradT)$. The composite material is described by effective coefficients, which by definition relate the volume averaged fields and flows, in this case $\langle \mathbf{e} \rangle$, $\langle gradT \rangle$, and $\langle \mathbf{j} \rangle$:

$$\langle \mathbf{j} \rangle = \sigma^{eff} \langle \mathbf{e} \rangle + \sigma^{eff} \alpha^{eff} \langle -gradT \rangle. \tag{A1}$$

It is important to note that, firstly, the effective coefficients, in particular the effective conductivity depends not only on the conductivity of the phases (in the case of two phases, $\sigma_1$ and $\sigma_2$) but on the thermal conductivity and the thermoelectric power values as well. Therefore, the effective factors are renormalized, including not only the values of local conductivity coefficients, but also the thermoelectric power and thermal conductivity. Secondly, even without the second term, when the "pure" conductivity problem is considered, the symmetries factor of the effective coefficient and local coefficients may be different. For example, a macroscopically inhomogeneous medium consisting of two phases with isotropic conductivity may have anisotropic effective conductivity $\boldsymbol{\sigma}^{eff}$.

From (A1) it can be clearly seen that the total resistance $R = \dfrac{1}{\sigma^{eff}} \dfrac{L}{S}$, specifying the effective conductivity $\sigma^{eff}$ is no longer the ratio of the current passing through the sample to the potential difference (voltage). This happens because of the contribution to the current from the second thermoelectric term.

### Acknowledgement

MS acknowledges financial support by OTH Regensburg in in the framework of cluster funding.

# REFERENCES

1. G. Filipcsei, I. Csetneki, A. Szilagyi, and M. Zrinyi, *Adv. Polym. Sci.* **206**, 137 (2007).

2. Y. Li, J. Li, W. Li, and H. Du, *Smart Mater. Struct.* **23**, 123001 (2014).

3. Ubaidillah, J. Sutrisno, A. Purwanto, and S. A. Mazlan, *Adv. Eng. Mater.* **17**, 563(2015).

4. A. M. Menzel, *Phys. Reps.* **554**, 1 (2015).

5. S. Odenbach, *Arch. Appl. Mech.* **86**, 269 (2016).

6. M. T. Lopez-Lopez, J. D. G. Durán, L. Yu. Iskakova, and A. Yu. Zubarev, *J. Nanofluids* **5**, 479 (2016).

7. S. Bednarek, *J. Magn. Magn. Mater*. **301**, 200 (2006).

8. X. Guan, X. Dong, and J. Ou, *J. Magn. Magn. Mater*. **320**, 158 (2008).

9. K. Danas, S.V. Kankanala, and N. Triantafyllidis, *J. Mech. Phys. Solids*, **60**, 120 (2012).

10. E. Galipeau, P. Ponte Castañeda, *Int. J. Solid Struct.* **49**, 17 (2012).

11. E. Galipeau, P. Ponte Castañeda, *Proc. R. Soc. A* **469**, 20130385 (2013).

12. G. Stepanov, E. Y. Kramarenko, and D. Semerenko, *J. Phys.: Conf. Ser.* **412**, 012031 (2013).

13. E. Callen and H. Callen, *Phys. Rev.* **139**, 455 (1965).

14. P. Morin, D. Schmitt, and E. T. Lacheisserie, *Phys. Rev. B* **21**, 1742 (1980).

15. V. M. Kalita, A. F. Lozenko, and S. M. Ryabchenko, *Low Temp. Phys.* **26**, 489 (2000).

16. V. M. Kalita, A. F. Lozenko, S. M. Ryabchenko, and P. A. Trotsenko, *Low Temp. Phys.* **31**, 794 (2005).

17**.** V. M. Kalita, I. Ivanova, and V. M. Loktev, *Phys. Rev. B* **78**, 104415 (2008).

18. V.M. Kalita, A.A. Snarskii, D. Zorinets, and M. Shamonin, *Phys. Rev. E* **93**, 062503 (2016).

19. A. Zubarev, *Phys. A Stat. Mech. Appl.* **392**, 4824 (2013).

20. A. Zubarev and D. Yu. Borin. *J. Magn. Magn. Mater.* **377**, 373 (2015).

21. G. Diguet, E. Beaugnon, and J. Y. Cavaillé, *J. Magn. Magn. Mater.* **321**, 396 (2009).

22. O. V. Stolbov,Y. L. Raikher, and M. Balasoiu, *Soft Matter* **7**, 8484 (2011).

23. Y. Han, A. Mohla, X. Huang, W. Hong, and L. E. Faidley, *Int. J. Appl. Mech.* **07**, 1550001 (2015).

24. H.-N. An, S. J. Picken, and E. Mendes, *Soft Matter* **8**, 11995 (2012).

25. G. V. Stepanov, D. Yu. Borin, Yu. L. Raikher, P. V. Melenev, and N. S. Perov, *J. Phys.: Condens. Matter* **20,** 204121 (2008)

26 J. M. Linke, D. Yu. Borin, and S. Odenbach, *RSC Adv*. **6,** 100407 (2016)

27. R. Weeber, S. Kantorovich, and C. Holm, *Soft Matter* **8**, 9923 (2012).

28. R.Weeber, S. Kantorovich, and C. Holm, *J. Magn. Magn. Mater.* **383**, 262 (2015).

29. R. Weeber, S. Kantorovich, and C. Holm, *J. Chem. Phys.* **143**, 154901 (2015).

30. G. Pessot, H. Löwen, and A. M. Menzel, *J. Chem. Phys.* **145**, 104904 (2016) .

31. P. Metsch, K. A. Kalina, C. Spieler, M. Kästner, *Comp. Mater. Sci.* **124**, 364–374 (2016).

32. E. Jarkova, H. Pleiner, H.-W. Müller, and H. R. Brand, *Phys. Rev. E* **68**, 041706 (2003).

33. S. Bohlius, H. R. Brand, and H. Pleiner, *Phys. Rev. E* **70**, 061411 (2004).

34. A. Dorfmann, R. Ogden, *Eur. J. Mech. A Solids* **22**, 497-507 (2003).

35. I. Brigadnov, A. Dorfmann, *Int. J. Solids Struct.* **40**, 4659-4674 (2003).

36. D. Ivaneyko, V. Toshchevikov, M. Saphiannikova, and G. Heinrich. *Soft Matter* **10**, 2213 (2014).

37. M. R. Dudek, B. Grabiec, and K.W.Wojciechowski, *Rev. Adv. Mater. Sci.* **14**, 167 (2007).

38. D. S. Wood and P. J. Camp, *Phys. Rev. E* **83**, 011402 (2011).

39. M. A. Annunziata, A.M. Menzel, H. Löwen, *J. Chem. Phys.* **138**, 204906 (2013).

40. G. Pessot, P. Cremer, D. Y. Borin, S. Odenbach, H. Löwen, and A. M. Menzel, *J. Chem. Phys.* **141**, 124904 (2014).

41. M. Tarama, P. Cremer, D. Y. Borin, S. Odenbach, H. Löwen, and A. M. Menzel, *Phys. Rev. E* **90**, 042311 (2014).

42. P. A. Sánchez, J. J. Cerdá, T. Sintes, and C. Holm, *J. Chem. Phys.* **139**, 044904 (2013).

43. J. J. Cerdá, P. A. Sánchez, C. Holm, and T. Sintes, *Soft Matter* **9**, 7185 (2013).

44. D. Ivaneyko, V. P. Toshchevikov, and M. Saphiannikova, *Soft Matter* **11**, 7627-7638 (2015).

45. B. F. Spencer, S. J. Dyke, M. K. Sain, and J. D. Carlson, *J. Eng. Mech.* **123**, 230-238 (1997).

46. O. Stolbov, Y. L. Raikher, G. Stepanov, A. Chertovich, E. Y. Kramarenko, and A. Khokhlov, *Polym. Sci., Ser. A* **52**, 1344-1354 (2010).

47. W. Li, Y. Zhou, and T. Tian, *Rheol. Acta*, **49**, 733-740 (2010).

48. L. Chen and S. Jerrams, *J. Appl. Phys.* **110**, 013513 (2011).

49. J.-T. Zhu, Z.-D. Xu, and Y.-Q. Guo, *Smart Mater. Struct.* **21**, 075034 (2011).

50. F. Guo, C.-B. Du, and R.-P. Li, *Adv. Mech. Eng.* **6**, 629386 (2014).

51. T. A. Nadzharyan, V. V. Sorokin, G. V. Stepanov, A. N. Bogolyubov, and E. Y. Kramarenko, *Polymer* **92**, 179-188 (2016).

52. S.-Y. Fu, X.-Q. Feng, B. Lauke, and Y.-W. Mai, *Compos. Part B Eng.* **39**, 933 (2008).

53. S. Torquato, *Random Heterogeneous Materials. Microstructure and Macroscopic Properties* (Springer Verlag, New York, 2002).

54. A.V. Ryzhkov, P.V. Melenev, M. Balasoiu, and Y.L. Raikher, *J. Chem. Phys.* **145**, 074905 (2016).

55. P. Creemer, H. Löwen, and A.M. Menzel, *Phys. Chem. Chem. Phys.* **18,** 26670-26690 (2016).

56. H. Sahin, X. Wang, X., and F. Gordaninejad, *J. Intell. Mater. Syst. Struct.* **20** , 2215 (2009).

57. J. Yang, X. Gong, H. Deng, L. Qin, and S. Xuan, *Smart Mater. Struct.* **21**, 125015 (2012).

58. X. Dong, N. Ma, M. Qi, J. Li, R. Chen, and J. Ou, *Smart Mater. Struct*. **21**, 075014 (2012).

59. I. A. Belyaeva, E.Yu. Kramarenko, G. V. Stepanov, V.V. Sorokin, D. Stadler, and M. Shamonin, *Soft Matter* **12**, 2901-2913 (2016).

60. Z. Hashin, *J. Appl. Mech.* **50**, 481 (1983).

61. R. M. Christensen, *Mechanics of Composite Materials* (Wiley, New York, 1979).

62. L. D. Landau, L. P. Pitaevskii, and E. M. Lifshitz, *Electrodynamics of Continuous Media*, 2nd ed., Course of Theoretical Physics Vol. 8 (Elsevier, Oxford, UK, 1984).

63. V. A. Buryachenko, *Micromechanics of Heterogeneous Materials* (Springer Verlag, New York, 2007).

64. A.F. Kabychenkov and F. V. Lisiovskii, *J. Exp. Theor. Phys.* **123**, 254 (2016).

65. E.Z. Melikhov and R.M. Farzetdinova, *J. Exp. Theor. Phys*. **122**, 1038 (2016).

66. H. An, S. J. Picken, and E. Mendes, *Polymer* **53**, 4164-4170 (2012).

67. V. V. Sorokin, E. Ecker, G. V. Stepanov, M. Shamonin, G. J. Monkman, E. Y. Kramarenko, and A. R. Khokhlov, *Soft Matter* **10**, 8765-76, (2014).

68. A. A. Snarskii, I. V. Bezsudnov, V. A. Sevryukov, A. Morozovskiy, and J. Malinsky, *Transport Processes in Macroscopically Disordered Media. From Mean Field Theory to Percolation* (Springer Verlag, New York, 2016).

69. H. Böse and R. Röder, *J. Phys.: Conf. Ser.* **149**, 012090 (2009).

70. H. See and R. Tanner, *Rheol. Acta* **42**, 166-170 (2003).

71. S. Abramchuk, E. Kramarenko, G. Stepanov, L. V. Nikitin, G. Filipcsei, A. R. Khokhlov, and M. Zrinyi, *Polym. Adv. Technol.* **18**, 883–890 (2007).

72. B. T. Borbath, S. Günther, D. Yu. Borin, Th. Gundermann, and S. Odenbach, *Smart Mater. Struct.* **21**, 105018 (2012).

73. D. Romeis, V. P. Toshchevikov and M. Saphiannikova, *Soft Matter*, Accepted Manuscript, DOI: 10.1039/C6SM01798C (2016).

74. D. Stauffer, A. Aharoni, *Introduction To Percolation Theory*, Revised 2$^{nd}$ Ed. (Taylor & Francis, London, 1994).

75. H.E. Stanley, *Introduction to Phase Transitions and Critical Phenomena* (*International Series of Monographs on Physics*), Revised Ed. (Oxford University Press USA, New York, 1971).